\documentclass[prd,showpacs,superscriptaddress,amsmath,amssymb,nofootinbib]{revtex4-2}
\usepackage{graphicx,epsf}% Include figure files
\usepackage{dcolumn}% Align table columns on decimal point
\usepackage{bm}% bold math

\usepackage{color,comment}
\def\red{\color{red}}

\newcommand{\bea}{\begin{eqnarray}}
\newcommand{\eea}{\end{eqnarray}}

\newcommand{\hide}[1]{}

\begin{document}

\title{Phase Structure of Quantum Improved Schwarzschild-(Anti)de Sitter Black Holes}

\author{Chiang-Mei Chen} \email{cmchen@phy.ncu.edu.tw}
\affiliation{Department of Physics, National Central University, Zhongli, Taoyuan 320317, Taiwan}
\affiliation{Center for High Energy and High Field Physics (CHiP), National Central University,
 Zhongli, Taoyuan 320317, Taiwan}

\author{Yi Chen} \email{yi592401@gmail.com}
\affiliation{Department of Physics, National Central University, Zhongli, Taoyuan 320317, Taiwan}

\author{Akihiro Ishibashi} \email{akihiro@phys.kindai.ac.jp}
\affiliation{Department of Physics, Kindai University, Higashi-Osaka, Osaka 577-8502, Japan}
\affiliation{Research Institute for Science and Technology, Kindai University, Higashi-Osaka, Osaka 577-8502, Japan}

\author{Nobuyoshi Ohta} \email{ohtan@ncu.edu.tw}
\affiliation{Department of Physics, National Central University, Zhongli, Taoyuan 320317, Taiwan}
\affiliation{Research Institute for Science and Technology, Kindai University, Higashi-Osaka, Osaka 577-8502, Japan}

\date{\today}

\begin{abstract}
We study the phase structure of quantum improved Schwarzschild-(A)dS black holes in asymptotically safe gravity.
Our results confirm some of the well-known properties of quantum black holes. For example, the quantum effect
provides a repulsive force in the core region near singularity which stabilizes the thermodynamically unstable
small black holes, and also creates a zero temperature state with finite size.
We suggest that this could be a candidate for dark matter. We find
a new second order phase transition between small and large
black holes for quantum improved Schwarzschild-Anti de Sitter black holes. We also discuss the black holes
with different spatial topologies and find a notable duality.
\end{abstract}

%\pacs{04.62.+v, 04.70.Dy, 04.50.Gh, 04.40.Nr}

\maketitle

%%%%%%%%%%%%%%%%%%%%%%%%%%%%%%%%%%%%%%%%%%%%%%%%%%%%%%%%%%%%%%%%%%%%%%
\section{Introduction}
%%%%%%%%%%%%%%%%%%%%%%%%%%%%%%%%%%%%%%%%%%%%%%%%%%%%%%%%%%%%%%%%%%%%%%

The black holes are known to have thermodynamic properties including a characteristic temperature and an
intrinsic entropy~\cite{Wald:1999vt}.
The thermodynamics of Schwarzschild black holes encounter unstable phase with negative heat capacity.
The phase structure becomes more fascinating when we take into account a negative cosmological constant.
For such Schwarzschild-Anti de Sitter (SAdS) black holes, the temperature has a minimal value which corresponds
to the divergent point of heat capacity, giving the border line of unstable small black holes with negative
heat capacity and stable large black holes with positive heat capacity. In particular, there is a first order
Hawking-Page (HP) transition~\cite{Hawking:1982dh} between large (stable) SAdS black holes and thermal
Anti de Sitter (AdS) at the critical temperature where both the black hole and thermal AdS have the same free energy.
Above this critical temperature, the free energy of a large black hole is lower than the free energy
of thermal AdS of equal temperature, causing the decay of thermal AdS into black holes.
These thermodynamic aspects of black holes were originally discovered by using quantum field theory on
a classical background spacetime, or by applying euclidean quantum gravity approach initiated in~\cite{Gibbons:1976ue}.
No quantum effects of gravity are considered in these results.
It is thus interesting to revisit these issues in black hole thermodynamics, in particular
the phase structure, taking into account recent advances in quantum theory of gravity.

Quantum effects of gravity are, in general, expected to play a significant role in the core region near
the black hole singularity.
This is indeed the case for asymptotically safe quantum gravity~\cite{Reuter:1996cp, perbook, rsbook,
Niedermaier:2006wt, Niedermaier:2006ns}.
This approach considers the possibility of having an interacting quantum field theory of gravitation,
originated from a non-Gaussian ultraviolet (UV) fixed point in the theory space,
in the nonperturbative renormalization group (RG) framework.
The existence of such a fixed point would permit having an RG trajectory in the theory space
flowing to it and characterized by all the dimensionless couplings remaining finite when the UV cutoff is removed.
In this way, the UV behavior of the theory becomes under control, and this provides a promising way to quantum gravity.
An important result in this case is that the coupling ``constants'' depend on the energy scale along the RG flow.
The essential issue in this approach is a suitable choice of identification of the energy scale with
some length scale in the considered solutions. We call this quantum improvement. This allows us to understand
how the quantum effects manifest themselves in the black hole geometry. Given a suitable identification,
it turns out that quantum effects provide a ``repulsive'' force at the core of black hole which stabilizes
the unstable phase of small black holes~\cite{Bonanno:2000ep, Bonanno:2006eu}.
For discussions of black holes in this context,
see~\cite{Falls:2010he, Koch:2014cqa, Eichhorn:2022bgu, Platania:2023srt} and references therein.

The main purpose of this paper is to study the thermodynamic phase structure of quantum improved static black
holes in the asymptotically safe quantum gravity. As the mass decreases, the temperature of small AdS black holes
reduces to zero, instead of rapidly increasing to infinity. Consequently, as we will show, the quantum effects
create a local maximal in all asymptotically flat and (A)dS black holes. The small black holes become stable and,
therefore there exist finite size zero temperature remnants.
We would like to suggest that the Planck size cold remnants could be a candidate for dark
matter.\footnote{
The possibility of ``cold remnants'' was already pointed out in Ref.~\cite{Bonanno:2006eu}.
In the context of the primordial black holes (PBHs) as dark matter (see, e.g.,
Refs.~\cite{Carr:2021bzv, Green:2020jor, Harada:2013epa} and references therein), only those with initial mass
larger than $10^{15}$g have so far been considered, since any PBHs with smaller mass $\lesssim 10^{15}$g
would have already evaporated away by now. The asymptotic safe quantum improvement provides an interesting mechanism
that can make even much smaller PBHs free from Hawking evaporation, and thus create stable black hole remnants.
It is interesting to study the astrophysical implications of such stable Planck size black hole remnants.
In the asymptotic safety there are a number of papers arguing that the gravitational effects reduce
the strength of interactions of fermions and scalars (not black hole remnants), thereby making them candidates
of dark matter, see, e.g. Refs.~\cite{Sannino:2014lxa, Eichhorn:2018vah}.
} % end of footnote
Apart from that, it also raises another interesting question: {\it Is there a new phase transition,
analogous to the HP transition, between stable quantum improved small black holes and thermal AdS}?
In order to clarify this problem, it is necessary to study the phase structure of quantum improved black holes.
We will find rich structure in the case of AdS spacetime. Note that the importance of AdS black holes has been
well recognized for some time in the context of AdS/CFT correspondence~\cite{Maldacena}, and also HP-type of
phase transition, so it is interesting to see how the results are modified by quantum effects~\cite{Ferrero:2022dpk}.

Here we address these questions and find that for the quantum improved SAdS black holes, when the quantum effects
are not significantly large, there is an intermediate unstable phase between stable small and large black holes.
Such kind of sandwich structure indicates a possibility of the existence of a second order phase transition from
small to large black holes or vice versa. The evidence of such a phase transition is the presence of
a swallow-tail pattern in free energy with respect to temperature.
The phase structure of a thermodynamical system is encoded in the associated free energy. Unfortunately due to
the lack of quantum action, we are not able to compute the free energy via the partition function.
However, by assuming the thermodynamic free energy is still valid for quantum improved black holes,
we can study the phase structure of quantum improved SAdS black holes. We find that the analogous HP
transition can indeed occur only at unphysical negative temperature. Moreover, a new interesting swallow-tail
pattern does appear when the strength of quantum effects are smaller than the critical value and the associated
second order phase transition can happen.

For the quantum improved asymptotically flat black holes, the heat capacity is independent of their spatial
topologies, and thus they all share similar phase structures.
When the cosmological constant is non-vanishing, we find that the heat capacity is invariant under sign change
of the cosmological constant $\Lambda \to - \Lambda$ and the spatial topologies $\sigma \to -\sigma$.
But, both the mass and temperature flip their signs. This means, for example, the phase structure of
topological Schwarzschild-de Sitter (SdS) ``black holes'' is alike those of SAdS black holes.
However, for the topological SdS black holes the mass is negative~\cite{Mann:1997jb}. The solutions for
$\sigma \ne 1$ and $\Lambda \ge 0$ correspond to cosmological solutions with positive mass parameter.

This paper is organized as follows. In the next section, we will briefly describe how to quantum improved black hole
geometries in the asymptotic safety scenario, and discuss the first law of a quantum improved Kerr-(A)dS black hole
as an example. Then, in Sec.~\ref{sec:3}, we will analyze quantum effects on thermodynamic phase structure of
asymptotically AdS static black holes with different horizon topologies. We compute the Hawking temperature,
heat capacity, and free energy for quantum improved AdS black holes, and for each type of horizon topology,
we study in detail whether quantum improved AdS black holes exhibit the HP type phase transition.
We find that for the spherical horizon case, a new type of second order phase transition can occur,
due to the effects of quantum improvement, whereas for the hyperbolic and the planar horizon case,
there are no phase transition  analogous to the HP. We perform similar analyses of
quantum improved black holes with positive cosmological constant in sec.~\ref{sec:4} and with vanishing
cosmological constant in sec.~\ref{sec:5}. Section~\ref{sec:6} is devoted to summary and discussions.
In Appendix~\ref{FEHP}, we briefly summarize how to compute free energy for SAdS black holes
in general relativity.

%%%%%%%%%%%%%%%%%%%%%%%%%%%%%%%%%%%%%%%%%%%%%%%%%%%%%%%%%%%%%%%%%%%%%
\section{Running Newton Coupling in Asymptotically Safe Gravity}
\label{sec:2}
%%%%%%%%%%%%%%%%%%%%%%%%%%%%%%%%%%%%%%%%%%%%%%%%%%%%%%%%%%%%%%%%%%%%%%

Let us first summarize some relevant results here in the asymptotically safe scenario for quantum generalization
of the general relativity with cosmological constant, in the units $c = \hbar = 1, G_0 = l_P^2 = 1/M_P^2$.
The action is given by
\begin{equation}
S = - \frac1{16 \pi G} \int d^4x \sqrt{-g} \left( R - 2 \Lambda \right).
\end{equation}
The asymptotically safe scenario proposes the Newton coupling $G(k)$ and cosmological ``constant'' $\Lambda(k)$
are energy dependent~\cite{Reuter:1996cp}.
Assuming the cosmological constant is is sufficiently small,
the associated renormalization group equations lead to the solutions for
Newton coupling~\cite{Bonanno:2000ep, Reuter:2010xb, Harst:2011zx, Pawlowski:2018swz, Ishibashi:2021kmf}
\begin{equation}
G(k) = \frac{G_0}{1 + \omega G_0 k^2},
\end{equation}
where $\omega > 0$ is the inverse of the fixed point value.
Actually the ``dimensionful'' cosmological constant $\Lambda$ is given by~\cite{Pawlowski:2018swz, Koch:2014cqa}
\bea
\Lambda = \Lambda_0 + \lambda_* k^2,
\label{cospara}
\eea
where $\lambda_*$ is the fixed point value of the ``dimensionless'' cosmological constant and
$\Lambda_0$ is the infrared value of the cosmological constant. It might appear that $\Lambda$ increases
for larger $k$ or near the singularity. However we are only concerned with the thermodynamic behaviors of
the black holes and the near-horizon area is the region we consider. The typical value of $\lambda*$ is
very small like $0.1$~\cite{CPR}. So our approximation can be justified.

How to identify the energy scale $k$ with a spacetime distance scale is an essential issue for constructing
quantum improved solutions~\cite{Reuter:2010xb}.
See also related work~\cite{Koch1, Koch2, Koch:2013owa, Koch:2015nva, Gonzalez:2015upa, Borissova:2022mgd}.
The first law of black hole thermodynamics, particularly for Kerr-(A)dS black holes, requires that a consistent
identification near the horizon should be a function of the horizon area $A_h$~\cite{Chen:2022xjk}.
The same consequence can be derived with the relation of entropy variation
$\delta S = \delta A_h/4 G$~\cite{Falls:2012nd} in which it is implicitly presumed from the outset that
the running coupling $G$ is a function of the horizon area $A_h$. Therefore, a physically admissible
identification for energy scale is a function of area with given radius. According to dimensional counting,
a naturally suggested scale identification for Kerr-(A)dS black holes at horizon is
\begin{equation}
k(r_h) = \xi \sqrt{\frac{1 + \Lambda a^2/3}{r_h^2 + a^2}},
\label{id}
\end{equation}
with a dimensionless parameter $\xi$. Strictly speaking, we may have $r_h$-dependence in the cosmological
constant, but we neglect this because we are interested only in the case of small cosmological constant.
This identification is the simplest and natural one without introducing additional dimensional parameters.
The resulting running Newton coupling is
\begin{equation}
\label{eq_Grha}
G(r_h) = \frac{G_0 (r_h^2 + a^2)}{r_h^2 + a^2 + \tilde\omega G_0 (1 + \Lambda a^2/3)}, \qquad
\tilde\omega = \xi^2 \omega.
\end{equation}
The black hole entropy can be computed by integrating the first law of thermodynamics which,
up to an integrating constant $S_0$, gives
\begin{equation}
S = \frac{\pi (r_h^2 + a^2)}{G_0 (1 + \Lambda a^2/3)} + \pi \tilde\omega
 \ln \frac{\pi (r_h^2 + a^2)}{G_0 (1 + \Lambda a^2/3)} + S_0.
\end{equation}
The temperature determined by the surface gravity $\kappa$ is
\begin{equation}
T_\mathrm{H} = \frac{\kappa}{2 \pi} = \frac{- \Lambda r_h^4 + \left[ 1 - \Lambda a^2/3
 - \tilde\omega G_0 (1 + \Lambda a^2/3) \Lambda/3 \right] r_h^2 - a^2
 - \tilde\omega G_0 (1 + \Lambda a^2/3)}{4 \pi r_h \left[ r_h^2 + a^2 + \tilde\omega G_0 (1 + \Lambda a^2/3) \right]}.
\label{HT}
\end{equation}
The quantum effects enlarge the size of extremal (zero temperature) state which can still exist in the static
black holes with finite mass. The stable extremal black holes could be a candidate of dark matter.
For non-rotating black holes $a = 0$, the vanishing condition of the temperature~\eqref{HT} gives
the extremal limit with
\begin{equation}
r_\mathrm{zero}^2 = \frac{1 - \tilde\omega G_0 \Lambda/3 \pm \sqrt{(1 - \tilde\omega G_0 \Lambda/3)^2
 - 4 \tilde\omega G_0 \Lambda}}{2 \Lambda},
\label{zerot}
\end{equation}
where $\pm$ corresponds to dS ($\Lambda > 0$)/AdS ($\Lambda < 0$) and $r_\mathrm{zero}^2 = \tilde\omega G_0$
for $\Lambda = 0$.

%%%%%%%%%%%%%%%%%%%%%%%%%%%%%%%%%%%%%%%%%%%%%%%%%%%%%%%%%%%%%%%%%%%%%%
\section{Asymptotically Anti-de Sitter Black Holes}
\label{sec:3}
%%%%%%%%%%%%%%%%%%%%%%%%%%%%%%%%%%%%%%%%%%%%%%%%%%%%%%%%%%%%%%%%%%%%%%

In this section we are going to analyze the quantum effects on the thermodynamical phase structure of
the asymptotically AdS black holes for $a=0$ with different horizon topologies, characterized by
the unit sectional curvature $\sigma$ of the horizon manifold, i.e. sphere ($\sigma = 1$), flat ($\sigma = 0$)
and hyperbola ($\sigma = -1$):
\begin{equation}
ds^2 = - f(r) dt^2 + \frac{dr^2}{f(r)} + r^2 d\Omega_\sigma^2, \qquad
f(r) = \sigma - \frac{2 G(r) M}{r} + \frac{r^2}{L^2}.
\label{def:metric}
\end{equation}
(We also refer to the hyperbola ($\sigma=-1$) case as the {\it topological} black hole.)
We also denote the AdS radius-squared as
\bea
L^2 = - 3/\Lambda.
\eea
The identification~\eqref{id} and~\eqref{eq_Grha} for non-rotating black holes
($a=0$) reduces to
\begin{equation}
\label{eq_Grh}
k(r_h) =  \frac{\xi}{r_h}, \qquad
G(r_h) = \frac{G_0 r_h^2}{r_h^2 + \tilde\omega G_0}.
\end{equation}

Note that for the static black hole metric (\ref{def:metric}), the horizon area is given merely
in terms of the radial coordinate $r$, different from the Kerr-(A)dS case, and therefore any scale
identification of the type $k = \tilde\xi/r^p$ with $p>0$ examined in~\cite{Ishibashi:2021kmf}
(with suitable dimension for $\tilde\xi$) is consistent with the black hole first-law discussed
in~\cite{Chen:2022xjk}. Then, depending on the power $p$,
the quantum improved geometry near the center $r=0$ will differ.
In particular, for the identification~\eqref{eq_Grha} with $a=0$, for which $k \sim \xi/r$, the metric function
behaves $f(r) \sim \sigma - (2M/\tilde{\omega}) r+ O(r^2)$ near the center and therefore the resultant
quantum improved geometry admits a weak singularity at the center.
However, since the temperature and heat capacity are computed at the horizon and also the euclidean path-integral
for the free energy is performed outside the horizon as shown in Appendix A, the behavior of the quantum improved
geometry (i.e., whether it is regular or not) near the center does not appear to affect much
the evaluation of these thermodynamic quantities. For this reason, we will stick to the scale
identification~\eqref{eq_Grha}, which is the simplest and natural choice without any additional dimensional parameters,
in the rest of this paper, despite possible more general dependence.

The location of the horizon $f(r_h) = 0$ gives the mass as
\begin{equation}
\label{eq_M}
M = \frac{(\sigma + r_h^2/L^2) (r_h^2 + \tilde\omega G_0)}{2 G_0 r_h}.
\end{equation}
We see from \eqref{cospara} and \eqref{eq_Grh} that there is actually some dependence on the horizon radius $r_h$
in the cosmological constant, but this could be neglected for small $\lambda_*$ unless we consider
very small $r_h$ region.
Moreover the dependence on the horizon radius $r_h$ is smaller for larger $r_h$ in which we will be interested.

The mass parameter $M$ is always positive for both $\sigma = 1$ and $\sigma = 0$. However, for $\sigma = -1$
and the ``horizon'' radius smaller than AdS radius, i.e. $r_h < L$, the mass parameter is negative.
It is straightforward to compute the Hawking temperature
\begin{equation}
T_\mathrm{H} = \frac{3 r_h^4/L^2 + (\sigma + \tilde\omega G_0/L^2) r_h^2
- \sigma \tilde\omega G_0}{4 \pi r_h (r_h^2 + \tilde\omega G_0)}
= \frac{\sigma + 3 r_h^2/L^2}{4 \pi r_h} - \frac{\tilde\omega G_0 (\sigma + r_h^2/L^2)}{2 \pi r_h (r_h^2
 + \tilde\omega G_0)},
\label{eq_T}
\end{equation}
and the heat capacity
\begin{equation}
C = \frac{2 \pi (r_h^2 + \tilde\omega G_0)^2 \left[ 3 r_h^4 + (\sigma L^2 + \tilde\omega G_0) r_h^2
- \sigma L^2 \tilde\omega G_0 \right]}{G_0 \left[ 3 r_h^6 - (\sigma L^2 - 8 \tilde\omega G_0) r_h^4
+ \tilde\omega G_0 (4 \sigma L^2 + \tilde\omega G_0) r_h^2 + \sigma L^2 \tilde\omega^2 G_0^2 \right]}.
\label{eq_C}
\end{equation}
The heat capacity, temperature and mass parameter have an interesting duality under transformation
$L^2 \to - L^2$ and $\sigma \to - \sigma$
\begin{equation}
\label{eq_duality}
C \to C, \qquad T \to - T, \qquad M \to - M.
\end{equation}

For studying the black holes thermodynamics, it is convenient to use the dimensionless variable $\rho$
\begin{equation}
\rho \equiv \frac{r_h^2}{L^2} \ge 0,
\end{equation}
and dimensionless parameter $\zeta$
\begin{equation}
\zeta \equiv \frac{\tilde\omega G_0}{L^2} \ge 0.
\end{equation}
The divergent points of heat capacity play important role in studying the thermodynamic phase structure.
This happens when the denominator of Eq.~\eqref{eq_C} vanishes. This leads to a cubic equation
\begin{equation}
d_\sigma(\rho) := 3 \rho^3 - (\sigma - 8 \zeta) \rho^2 + \zeta (4 \sigma + \zeta) \rho + \sigma \zeta^2 = 0,
\label{divh}
\end{equation}
and its discriminant is
\begin{equation}
\label{Dsigma}
D_\sigma := 4 \zeta^2 (\zeta - \sigma) (13 \zeta^3 - 303 \sigma \zeta^2 + 327 \sigma^2 \zeta - 5 \sigma^3).
\end{equation}
In the following we will discuss the phase structure for black holes with different spatial topology.

\subsection{$\sigma = 1$}
%%%%%%%%%%%%%%%%%%%%%%%%%%%%%%%%%%%%%%%%%%%%%%%%%%%%%%%%%%%%%%%%%%%%%%

\begin{figure}
\includegraphics[scale=1, angle=0]{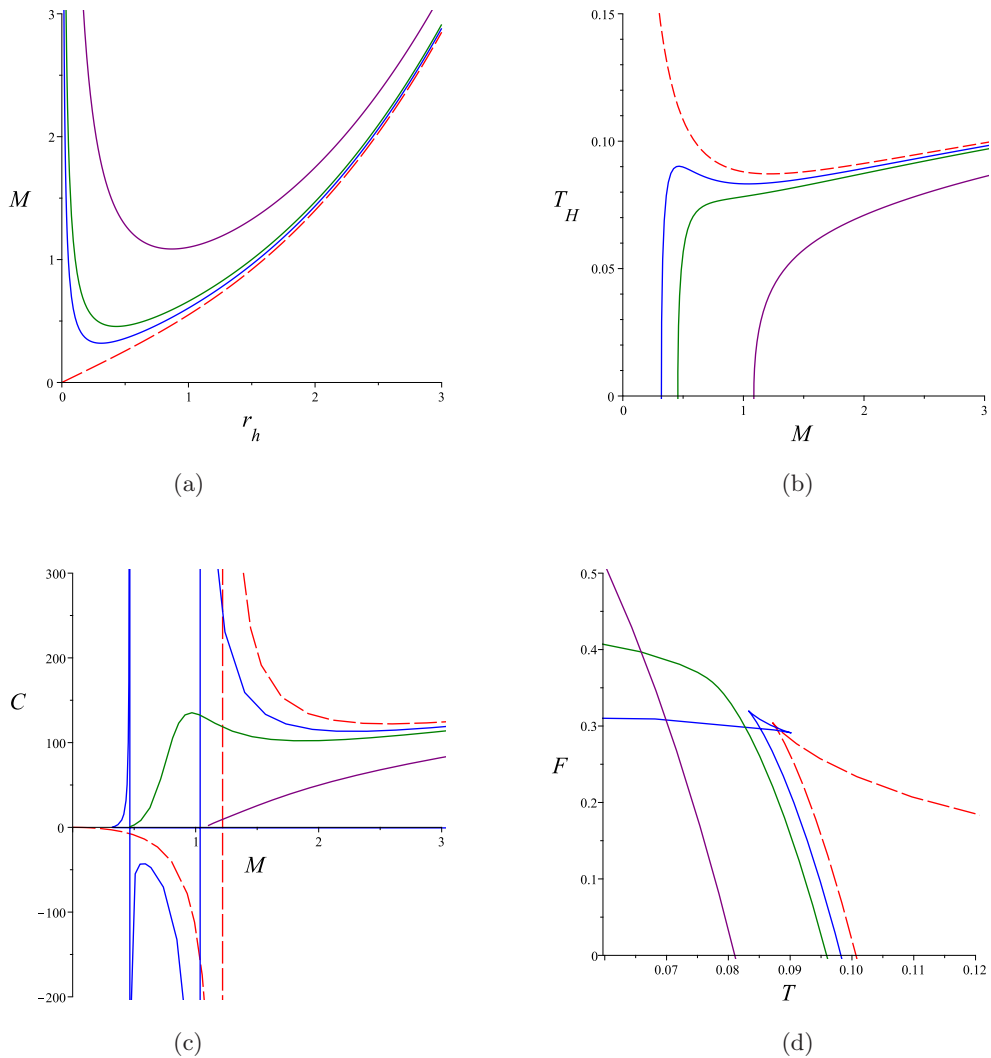}
\caption{(a) The horizon radius, (b) Hawking temperature, (c) heat capacity and (d) free energy of
quantum improved SAdS black holes with $G_0 = 1, L^2 = 10$. The red dash line represents the classic result
without quantum improvement. The blue solid line corresponds to $\tilde\omega = 0.1 \, (\zeta = 0.01)$ in which
there is a unstable phase with negative heat capacity. The green and purple solid lines correspond
to $\tilde\omega = 0.2, 1 \, (\zeta = 0.02, 0.1)$ in which the phase is all stable.}
\label{fig_AdSp}
\end{figure}

For the typical black holes with $\sigma = 1$, it is straightforward to plot the mass~\eqref{eq_M} with respect
to the horizon radius and the temperature~\eqref{eq_T} and the heat capacity~\eqref{eq_C} with respect to
the mass parameter.
These are given in Figs.~\ref{fig_AdSp}~(a)--(c) for different choice of $\tilde\omega$.
We note that these figures are drawn for the fixed cosmological constant $L^2 = 10$, i.e. $|\Lambda| = 0.3$.
Strictly speaking it depends on the horizon radius $r_h$.
However, since typically $\lambda_* \sim 0.1$, our approximation is valid at least for $r_h$ larger than 1,
and better for larger $r_h$. Similar remarks apply to other figures.

We see from Fig.~\ref{fig_AdSp}~(a) that the quantum improved black holes have
two horizons for the mass parameter greater than a lower bound which corresponds to extremal zero temperature
state $T_\mathrm{H} = 0$. The radius of zero temperature state is, from \eqref{zerot},
\begin{equation}
\label{eq_r0}
r_\mathrm{zero}^2 = \frac{\sqrt{L^4 + 14 \tilde\omega G_0 L^2 + \tilde\omega^2 G_0^2} - L^2 - \tilde\omega G_0}{6},
\end{equation}
and the associated entropy is
\begin{equation}
S_\mathrm{BH} = \frac{\pi r_\mathrm{zero}^2}{G_0} + \pi \tilde\omega \ln \frac{\pi r_\mathrm{zero}^2}{G_0}
 + \pi \tilde\omega S_0 = \frac{\pi r_\mathrm{zero}^2}{G_0} \left( 1 + \frac{\tilde\omega G_0}{r_\mathrm{zero}^2}
 \ln \frac{\pi r_\mathrm{zero}^2}{G_0} + \frac{\tilde\omega G_0}{r_\mathrm{zero}^2} S_0 \right).
\end{equation}
The logarithmic term is similar to the one obtained as quantum corrections~\cite{Kaul:2000kf}
and our result is reasonable for the quantum case.
Similar result is also obtained in loop quantum gravity~\cite{Meissner:2004ju}.

In Fig.~\ref{fig_AdSp}~(c), one can observe that the existence of $\tilde\omega$ creates a thermodynamically
``stable'' (not run away) phase with positive heat capacity in the small $M$ region. Consequently there are
two interesting results: (i) The quantum improved black holes can have a zero temperature state with
finite mass which could give a candidate for dark matter. (ii) Instead of the HP transition
between large black holes and AdS thermal states, quantum effects may create new kind, either first order
or second order, of phase transitions.

In GR, there is a first order HP phase transition in the SAdS black holes. It is interesting to
check how the quantum effects modify  the ``original'' phase transition or even generate new kinds of
phase transition. The number of divergent points of heat capacity is closely related to the phase structure
of thermodynamics. The thermodynamic phase structure is determined by the divergent points of
heat capacity~\eqref{divh}, i.e. the roots of the following cubic equation for dimensionless variable $\rho$:
\begin{equation}
d_+(\rho) := 3 \rho^3 - (1 - 8 \zeta) \rho^2 + \zeta (4 + \zeta) \rho + \zeta^2 = 0,
\label{divC}
\end{equation}
and its discriminant is
\begin{equation}
D_+ := 4 \zeta^2 (\zeta - 1) (13 \zeta^3 - 303 \zeta^2 + 327 \zeta - 5).
\label{Dp}
\end{equation}
There are 5 real roots of $D_+=0$, numerically given as
\begin{equation}
\zeta_0 = 0, \quad
\zeta_1 = 0.01551337272, \quad
\zeta_2 = 1, \quad
\zeta_3 = 1.118084244, \quad
\zeta_4 = 22.17409469.
\end{equation}

From the sign of the last term of Eq.~\eqref{divC}, we see that the product of the 3 roots is negative, so
Eq.~\eqref{divC} always has at least one negative root.
For $\zeta$ in the range $\zeta_0<\zeta<\zeta_1, \zeta_2<\zeta<\zeta_3$ and $\zeta_4<\zeta$,
the discriminant~\eqref{Dp} is positive, and we have 3 real roots of Eq.~\eqref{divC}.
From the derivative of Eq.~\eqref{divC}
\begin{equation}
d_+(\rho)' = 9 \rho^2 - 2(1 - 8 \zeta) \rho + \zeta (4 + \zeta),
\end{equation}
we see that this is positive for $\zeta>\frac18$ and positive $\rho$. This means that $d_+(\rho)$ monotonically
increases for positive $\rho$, and together with $d_+(0)>0$, there is no positive root for $\zeta>\frac18$.
For $\zeta_1<\zeta\leq\frac18$, where the discriminant~\eqref{Dp} is negative, there are two complex and one negative
roots. We then find that there are two positive roots for $\rho$ only when the value of parameter $\zeta$ is
in the region $0 < \zeta < \zeta_1$ (the region where the heat capacity is negative) and a degenerate positive root
at $\zeta = \zeta_1$ (region of negative heat capacity shrinking to a point).

When the quantum effects are small ($\zeta < \zeta_1$), the two positive roots of $\rho$, corresponding to
two real and positive roots of $r_h$, are the locations of local maximum and minimum of the temperature in accord
with the divergent points of heat capacity. In between the two divergent points, the thermal black hole system
is unstable with negative heat capacity.
This case is depicted by blue solid lines, together with other values of $\zeta$, in Figs.~\ref{fig_AdSp}.
The physical picture of SAdS black holes is the following:
In GR, the black hole temperature has a minimum corresponding to the divergent point separating negative
(unstable small black hole states) and positive (stable large black hole states) regions of heat capacity.
Moreover, there is a first order HP phase transition between a stable large black hole phase and
thermal AdS state. The quantum effects provide a ``repulsive'' force and generates stable phases in small mass regions.
This gives two possible new phase transitions: (i) a first order phase transition (analogy of the HP transition
in classical solutions) between small black holes and AdS thermal state, (ii) a second order phase transition
in between stable small and large black holes.

However, as the quantum effects become stronger, two divergent points become closer and degenerate
at the critical value $\zeta = \zeta_1$, and beyond it the heat capacity will become smooth and the corresponding
black hole thermal system is always stable (green solid lines in Fig.~\ref{fig_AdSp}).

In order to study the details of phase transition, we have to compute the free energy. In GR, as discussed
in Appendix~\ref{FEHP}, the difference in the free energies of SAdS black holes and AdS thermal state can be computed
either by saddle approximation of partition function or directly by thermodynamic relation $F = E - T S$.
Since we do not have quantum action for asymptotically safe gravity to compute the associated partition
function, we simply use the second relation $F = E - T S$ to calculate the difference in the free energies
of quantum improved SAdS black holes and thermal AdS state.

The result is depicted in Fig.~\ref{fig_AdSp}~(d) as a function of the temperature.
We find a swallow-tail pattern. Such a pattern exists within the range of the dimensionless parameter
$0 < \zeta < \zeta_1$ in which the temperature can have both the local maxima and minima.
A second order phase transition between the small and large black holes can
occur at the intersecting point. However, when $\zeta$ is larger than the critical value $\zeta_1$,
both the local maximal and minimal temperatures disappear and all phases are stable. A similar swallow-tail
pattern was observed in the asymptotically flat black holes~\cite{Mandal:2022quv} by using the identification
proposed in~\cite{Bonanno:2006eu}.

If the first order transition of the quantum improved small black holes and AdS thermal state were to occur,
it could occur only at unphysical negative temperature because the free energy is positive when temperature is
smaller than HP temperature and is still positive at $T = 0$ so the free energy can have
only one root (``original'' HP) in positive $T$ region.
Indeed, if such kind of phase transition occurred at $T > 0$, then the free energy should be negative
at zero temperature corresponding to $r_h = r_\mathrm{zero}$. However, from~\eqref{eq_M} and~\eqref{eq_r0},
it is obvious that  $F(r_h = r_\mathrm{zero}) = M(r_h = r_\mathrm{zero})$ is positive. Thus, no HP like
first order phase transitions (in addition to the original HP transition) could be generated for $T > 0$.
Mathematically, the free energy can be negative at both small and large $r_h$.
For small $r_h$ the free energy becomes negative only at negative temperature
region, i.e. $r_h < r_\mathrm{zero}$ which is unphysical. For large $r_h$ the quantum effects
``reduce'' the critical temperature of the ``classical'' HP transition, see also Fig.~\ref{fig_AdSp}~(d).
However, it is hard to find the explicit expression for critical temperature due to the logarithmic term
in the entropy.

\subsection{$\sigma = -1$}
%%%%%%%%%%%%%%%%%%%%%%%%%%%%%%%%%%%%%%%%%%%%%%%%%%%%%%%%%%%%%%%%%%%%%%

\begin{figure}
\includegraphics[scale=1, angle=0]{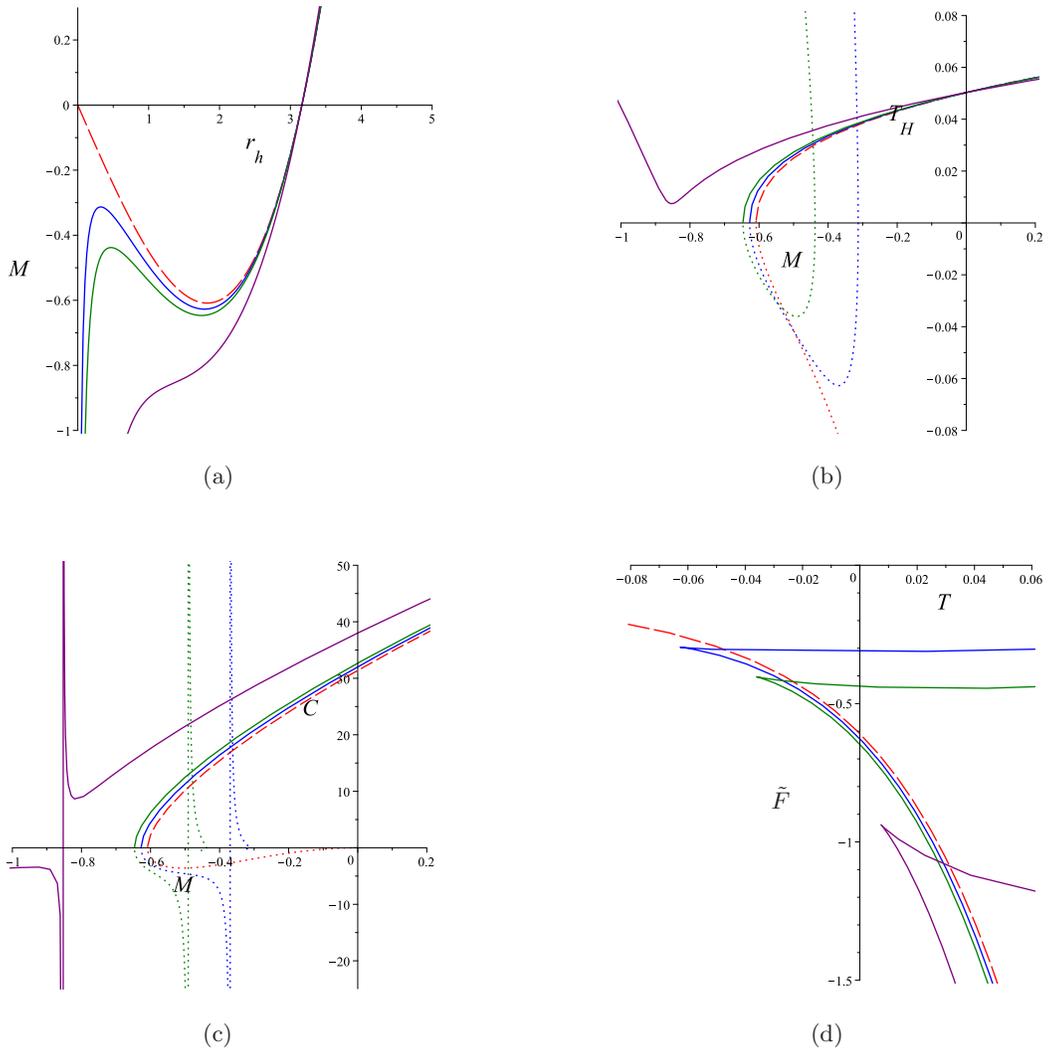}
\caption{(a) The horizon radius, (b) Hawking temperature, (c) heat capacity and (d) free energy density
for quantum improved ``topological SAdS black holes'' ($\sigma = -1$) with $G_0 = 1, L^2 = 10$.
The red dash line represents classic result without quantum improvement. The blue, green and purple solid lines
correspond to $\tilde\omega = 0.1, 0.2, 1 \, (\zeta = 0.01, 0.02, 0.1$).}
\label{fig_AdSn}
\end{figure}

For topological black holes ($\sigma = -1$) in GR without quantum effects ($\tilde\omega=0$), we depict
the mass~\eqref{eq_M} with respect to the horizon radius and the temperature~\eqref{eq_T} and heat
capacity~\eqref{eq_C} with respect to the mass parameter in Fig.~\ref{fig_AdSn} by red dashed lines.
We find that there is only one horizon for positive $M$, and two horizons for negative $M$ larger than
the lower bound, as can be seen from Fig.~\ref{fig_AdSn}~(a).
The extremal limit can be found by imposing the degenerate condition on $f(r)$ in~\eqref{def:metric} with
constant $G(r)=G_0$:
\begin{equation}
r_h = \frac{L}{\sqrt3}, \qquad
M = - \frac{L}{3 \sqrt3 G_0},
\end{equation}
which gives the lower bound of $M$ of black hole solutions. Here the temperature is zero.

Including the quantum effects, the allowed mass parameter is from $M = -\infty$ ($r_h = 0$) to $M = \infty$
($r_h = \infty$).
The temperature in Fig.~\ref{fig_AdSn}~(b) is plotted for full range of horizon radius $r_h$
in order to illustrate the duality~\eqref{eq_duality} by comparing with the plots of quantum
improved de Sitter black holes. Thus the dot parts in Fig.~\ref{fig_AdSn}~(b) and Fig.~\ref{fig_AdSn}~(c)
are the results of the ``inner'' horizon.

In GR, the topological AdS black holes are always stable. For the quantum improved black holes,
the present discriminant $D_-$ [obtained by setting $\sigma=-1$ in Eq.~\eqref{Dsigma}]
is always positive, so there are three real roots of $\rho$ for the equation $d_-(\rho)=0$.
Moreover, the derivative of $d_-(\rho)$ has two roots and the smaller one is always negative
which corresponds to a local maximum. Together with the fact that $d_-(-\infty) < 0, \, d_-(0) < 0$
and $d_-(\infty) > 0$, this means that the two roots must be negative and one is positive.
The heat capacity has only one divergent point which separates the stable and unstable phases.
The temperature has only a local minimum corresponding to the divergent point of heat capacity, as shown in
Fig.~\ref{fig_AdSn}~(b). If the quantum effect is small, the unstable phase is located inside the horizon,
but if the quantum effect is large enough, the small quantum improved black holes,
with more negative mass parameter, can be unstable.

For the topological AdS black holes, the hyperbolic surface with unit radius, defined by $x^2 + y^2 - z^2 = -1$,
has infinite area given by $\Omega_- = 2 \pi \int_0^\infty \sinh \nu d\nu$. The total energy and entropy
are infinite large quantities. Thus we should consider their ``densities'' as (the integration constant term
in the entropy vanishes because it is divided by the infinite area)
\begin{equation}
\tilde E = \frac{4 \pi}{\Omega_-} E = M = \frac{(-1 + r_h^2/L^2) (r_h^2 + \tilde\omega G_0)}{2 G_0 r_h}, \qquad
\tilde S = \frac{4 \pi}{\Omega_-} S = \frac{\pi r_h^2}{G_0} + \pi \tilde\omega \ln \frac{r_h^2}{4 G_0}.
\end{equation}
The last term is the logarithmic corrections from the quantum effects
and our result is again reasonable for the quantum case.
Then the free energy density is
\begin{equation}
\tilde F = \frac{4 \pi}{\Omega_-} F = \tilde E - T \tilde S.
\end{equation}
From the free energy density plotted in Fig.~\ref{fig_AdSn}~(d), we find that there is no analogous
HP transition.

%\begin{figure} 2-----
%\end{figure} 2----

\subsection{$\sigma = 0$}
%%%%%%%%%%%%%%%%%%%%%%%%%%%%%%%%%%%%%%%%%%%%%%%%%%%%%%%%%%%%%%%%%%%%%%

\begin{figure}
\includegraphics[scale=1, angle=0]{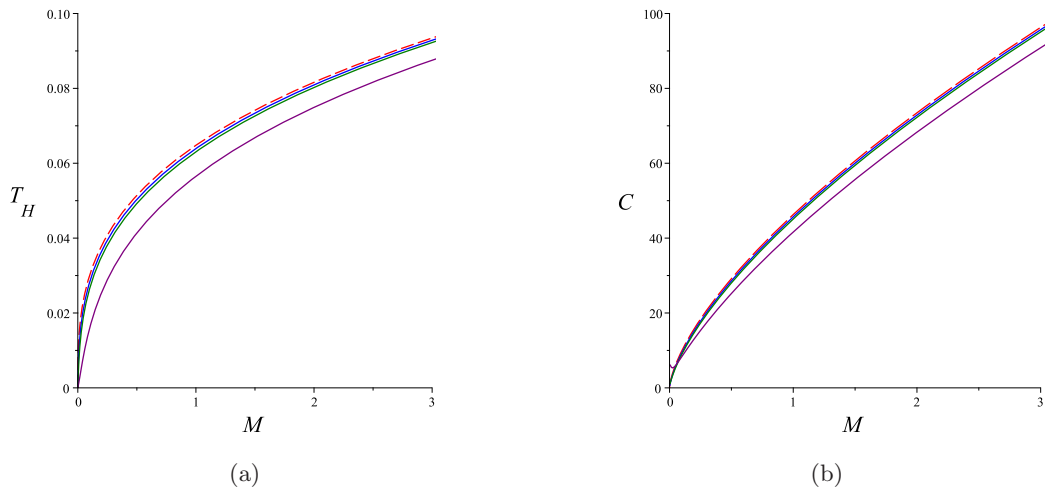}
\caption{(a) The Hawking temperature and (b) heat capacity for quantum improved ``planar SAdS black holes''
($\sigma = 0$) with $G_0 = 1, L^2 = 10$. The red dash line represents classic result without quantum
improvement. The blue, green and purple solid lines correspond to
$\tilde\omega = 0.1, 0.2, 1 \, (\zeta = 0.01, 0.02, 0.1$).}
\label{fig_AdS0}
\end{figure}

For the planar black holes ($\sigma = 0$), we plot the mass, temperature and heat capacity in Fig.~\ref{fig_AdS0}.
They are all positive, and the heat capacity is never divergent. Thus the quantum improved planar AdS black holes
do not change the stable nature in GR. The phase structure does not have an obvious change due to the quantum effects.

%\begin{figure} 3 ---
%\end{figure} 3 ---

%%%%%%%%%%%%%%%%%%%%%%%%%%%%%%%%%%%%%%%%%%%%%%%%%%%%%%%%%%%%%%%%%%%%%%
\section{Asymptotically de Sitter Black Holes}
\label{sec:4}
%%%%%%%%%%%%%%%%%%%%%%%%%%%%%%%%%%%%%%%%%%%%%%%%%%%%%%%%%%%%%%%%%%%%%%

The thermodynamic quantities for de Sitter black holes can be obtained simply from the corresponding
results in Anti-de Sitter cases by replacing $L^2 \to - L^2$, namely
\begin{eqnarray}
&& M = \frac{(\sigma - r_h^2/L^2) (r_h^2 + \tilde\omega G_0)}{2 G_0 r_h}, \qquad T_\mathrm{H}
 = \frac{- 3 r_h^4/L^2 + (\sigma - \tilde\omega G_0/L^2) r_h^2
 - \sigma \tilde\omega G_0}{4 \pi r_h (r_h^2 + \tilde\omega G_0)},
\nonumber\\
&& C = \frac{2 \pi (r_h^2 + \tilde\omega G_0)^2 \left[ 3 r_h^4 - (\sigma L^2 - \tilde\omega G_0) r_h^2
 + \sigma L^2 \tilde\omega G_0 \right]}{G_0 \left[ 3 r_h^6 + (\sigma L^2 + 8 \tilde\omega G_0) r_h^4
 - \tilde\omega G_0 (4 \sigma L^2 - \tilde\omega G_0) r_h^2 - \sigma L^2 \tilde\omega^2 G_0^2 \right]}.
\end{eqnarray}
Let us discuss the thermodynamic properties for three different $\sigma$ separately.

\subsection{$\sigma = 1$}
%%%%%%%%%%%%%%%%%%%%%%%%%%%%%%%%%%%%%%%%%%%%%%%%%%%%%%%%%%%%%%%%%%%%%%

According to the duality, it is not a surprise that the figures of thermodynamic quantities in Figs.~\ref{fig_dSp}
are a kind of ``reflection'' of the related quantities in Figs.~\ref{fig_AdSn} for topological AdS black holes.

In GR, the mass of the SdS black holes has an upper bound corresponding
to the degeneracy of the black hole and cosmological horizons. Only when the black holes horizon radius
is not larger than the radius of cosmological horizon, the Hawking temperature is positive and physical.

With quantum effects, there is another zero temperature state whose radius is given by~\eqref{eq_r0}
with $L^2$ replaced by $-L^2$. Only the black hole mass radius is in between this zero temperature radius
and cosmological radius, the Hawking temperature is positive, as shown by the solid lines in Fig.~\ref{fig_dSp}~(b).
If quantum effects are large enough, there are no solutions with positive Hawking temperature.

In GR, the de Sitter black holes is unstable with negative heat capacity. With the quantum effects,
the small black holes can be stable, see Fig.~\ref{fig_dSp}~(c). There are no phase transitions in de Sitter
black holes, as can be seen from Fig.~\ref{fig_dSp}~(d).

\subsection{$\sigma = -1$ and $\sigma = 0$}
%%%%%%%%%%%%%%%%%%%%%%%%%%%%%%%%%%%%%%%%%%%%%%%%%%%%%%%%%%%%%%%%%%%%%%

\begin{figure}
\includegraphics[scale=1, angle=0]{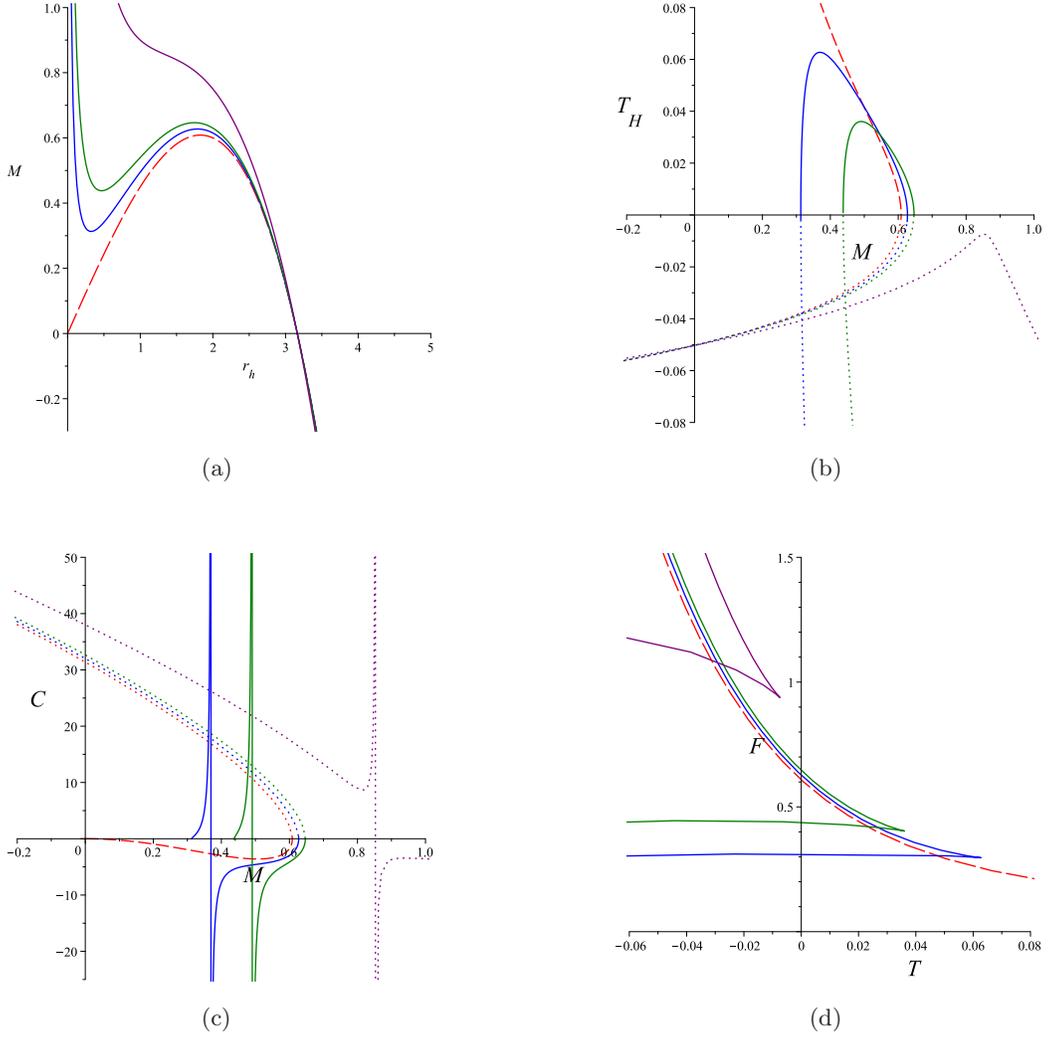}
\caption{(a) The horizon radius, (b) Hawking temperature, (c) heat capacity and (d) free energy for quantum improved
SdS black holes with $G_0 = 1, L^2 = 10$. The red dash line represents classic result without
quantum improvement. The blue, green and purple solid lines correspond to
$\tilde\omega = 0.1, 0.2, 1 \, (\zeta = 0.01, 0.02, 0.1$).}
\label{fig_dSp}
\end{figure}

For the cases $\sigma = -1$ and $\sigma = 0$ of asymptotically dS, there is no ``horizon'' within positive
value of $M$. Moreover, the lapse function
\begin{equation}
f(r) = \sigma - \frac{2 G(r) M}{r} - \frac{r^2}{L^2},
\end{equation}
is always negative. This means that the $r$ is actually the time coordinate in these cases which would give
cosmological solutions. Therefore, the black holes can exist if the mass parameter is negative.
However, $\sigma = -1$ solution is ``dual'' to anti-de Sitter black holes with $\sigma = 1$. From
Fig.~\ref{fig_dSp}, according to the duality~\eqref{eq_duality}, we see that the solutions with negative mass have
negative Hawking temperature.

%\begin{figure} 4---
%\end{figure} 4---

%%%%%%%%%%%%%%%%%%%%%%%%%%%%%%%%%%%%%%%%%%%%%%%%%%%%%%%%%%%%%%%%%%%%%%
\section{Asymptotically Flat Black Holes}
\label{sec:5}
%%%%%%%%%%%%%%%%%%%%%%%%%%%%%%%%%%%%%%%%%%%%%%%%%%%%%%%%%%%%%%%%%%%%%%
For this case, the heat capacity reduces to
\begin{equation}
C = - \frac{2 \pi (r_h^2 + \tilde\omega G_0)^2 (r_h^2 - \tilde\omega G_0)}{G_0 (r_h^4 - 4 \tilde\omega G_0 r_h^2
 - \tilde\omega^2 G_0^2)},
\end{equation}
which does not depend on the sign of $\sigma$. The divergent point of the heat capacity is then determined
by the equation:
\begin{equation}
r_h^4 - 4 \tilde\omega G_0 r_h^2 - \tilde\omega^2 G_0^2 = 0.
\end{equation}
The only positive root is $r_h^2 = (2 + \sqrt5) \tilde\omega G_0$.

\begin{figure}
\includegraphics[scale=1, angle=0]{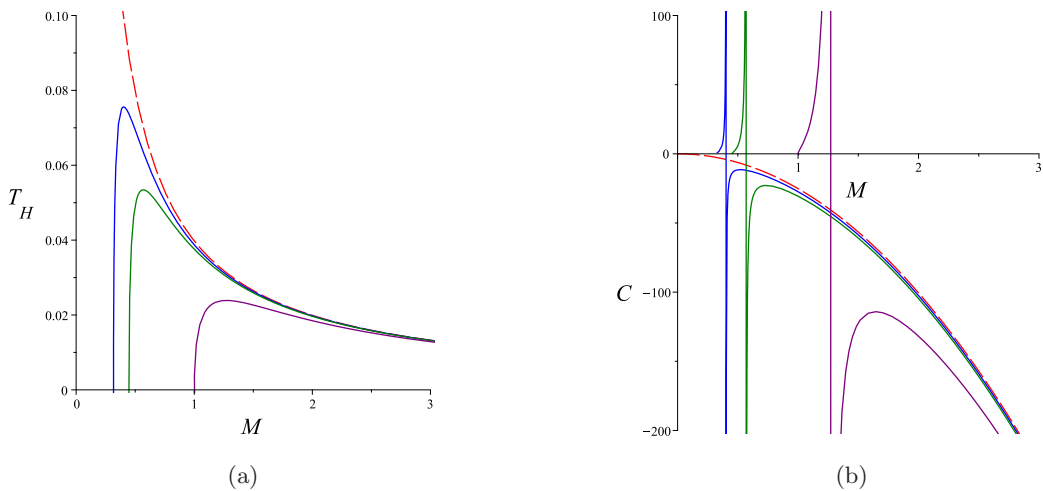}
\caption{ (a) The Hawking temperature and (b) heat capacity for quantum improved Schwarzschild black holes with
$G_0 = 1$. The red dash line represents classic result without quantum improvement. The blue, green and
purple solid lines correspond to $\tilde\omega = 0.1, 0.2, 1 \, (\zeta = 0.01, 0.02, 0.1$).}
\label{fig_Minp}
\end{figure}

The Schwarzschild black holes ($\sigma = 1$) are always unstable in GR, and Fig.~\ref{fig_Minp} shows that
the quantum effects stabilize the small black holes.
For $\sigma = -1$, the black holes solutions may exist for negative mass parameter.
However, according to the duality, and the Fig.~\ref{fig_Minp}~(a),
the associated Hawking temperature is negative.
For $\sigma = 0$, there are no black hole solutions.

This exhausts the quantum gravity effects in all possible cases.

%\begin{figure} 5 ---
%\end{figure} 5 ---

%%%%%%%%%%%%%%%%%%%%%%%%%%%%%%%%%%%%%%%%%%%%%%%%%%%%%%%%%%%%%%%%%%%%%%
\section{Conclusion}
\label{sec:6}
%%%%%%%%%%%%%%%%%%%%%%%%%%%%%%%%%%%%%%%%%%%%%%%%%%%%%%%%%%%%%%%%%%%%%%

We have studied the quantum effects of gravity on various black holes with and without cosmological constant.
Our findings can be summarized as follows.
\begin{itemize}
\item
For SAdS black holes, there always exists one and only one divergent point in GR
 (boundary of stable and unstable phases). Quantum improvement, with small quantum-effect parameter
$\omega$, increases one more divergent point and creates a new stable phase in the small mass
region (consequently generates a new zero temperature state with finite size). If $\omega$ increases,
the ``distance'' (in mass coordinate) between two divergent points reduces, and when $\omega$ is larger than
the critical value, both divergent points and unstable phase disappear,  see Fig.~\ref{fig_AdSp}~(b).
If $\omega$ is not too big, the black holes can have a second order phase transition.

\item
For the topological AdS black holes, the mass parameter can be negative (black holes with negative parameter
had been discussed in~\cite{Mann:1997jb, Hull:2022sgk}). The system is thermodynamically stable
if the quantum effects are small. Otherwise, the small black holes become unstable when the quantum effects
are big enough.

\item
For the planar AdS black holes ($\sigma = 0$), since the black holes is always stable in GR, the ``repulsive''
effects does not give an obvious change.

\item
For the SdS black holes, there are no divergent point in GR (only unstable phase exists).
The quantum improvement again creates a new stable phase in small mass region which then also generates
a divergent point. If the quantum improvement is big enough, there are no black hole solution with positive
Hawking temperature, see Fig.~\ref{fig_dSp}~(b).

\item
There is a duality in black hole thermodynamics. The heat capacity is invariant under the transformation
$L^2 \to -L^2$ and $\sigma \to -\sigma$, but the mass and temperature change sign. The topological
SdS black holes share various properties with SAdS black holes.

\end{itemize}

As listed above, we found that quantum improved Schwarzschild black holes with negative cosmological constant
have a richer phase structure than their classical counter-parts, as well as those with positive and vanishing
cosmological constant. For this, we have however assumed that the thermodynamic free energy is valid for the black
holes quantum improved at the solution level. It would be interesting to investigate whether one can also
compute the free energy by considering quantum improvement of the reduced action for euclidean black holes.
In this respect, once such an action for the quantum improvement is formulated, it would also be interesting to study the phase structure of quantum improved AdS black holes from the perspective of the AdS/CFT correspondence.

In the cosmological context, we are more concerned with quantum improved Schwarzschild black holes with positive
or vanishing cosmological constant. In general, primordial black holes of mass $\sim 10^{14}$g produced
in the early universe are expected to be undergoing now their final stages of the Hawking evaporation.
However, once the quantum effects are turned on, the evaporating process would stop at some point where
their horizon size is still finite, but the Hawking temperature is vanishing as can be seen in Fig.~\ref{fig_dSp}
and \ref{fig_Minp}. Thus, such primordial black holes eventually settle down to thermodynamically stable,
zero-temperature remnants, rather than being completely evaporated away. Although they may typically be
Planck-size objects, such remnants might possibly be a candidate of dark matter.
It is worth pursuing such an intriguing possibility.

In this paper, we have chosen the simplest identification~\eqref{id} to study the phase structure of the black holes.
In this case, we find that the singularity at the center of the black holes may not be resolved.
Unfortunately, in the Kerr black holes, the consistency of the thermodynamics
requires that the scale $k$ should be a function of the horizon area, and naive extension of this away from
the horizon does not allow such singularity resolution~\cite{Chen:2022xjk}.
However, it is also possible to consider the possibility to include the mass parameter of the black hole
in the identification. Indeed, with such dependence, it is possible that the singularity may be
resolved~\cite{Bonanno:2000ep}. It would be interesting to
investigate what restriction emerges when we allow the mass parameter in the identification,
and whether we may get the possibility that the singularity may be resolved.

\acknowledgments
The work of C.M.C. was supported by the National Science and Technology Council of the R.O.C. (Taiwan)
under the grants NSTC 111-2112-M-008-012.
The work of A.I. was supported in part by JSPS KAKENHI Grants No. 21H05182, 21H05186, 20K03938, 20K03975, 17K05451,
and 15K05092.
The work of N.O. was supported in part by the Grant-in-Aid for Scientific Research Fund of the JSPS (C)
No. 16K05331, 20K03980, and Taiwan NSTC 111-2811-M-008-024.

\begin{appendix}

%%%%%%%%%%%%%%%%%%%%%%%%%%%%%%%%%%%%%%%%%%%%%%%%%%%%%%%%%%%%%%%%%%%%%%
\section{Free Energy and Hawking-Page Transition}
\label{FEHP}
%%%%%%%%%%%%%%%%%%%%%%%%%%%%%%%%%%%%%%%%%%%%%%%%%%%%%%%%%%%%%%%%%%%%%%

For the classical SAdS black holes, the free energy can be computed from the action.
In $(d+1)$-dimensional spacetimes, the action is ($\Lambda = - d (d - 1)/2 L^2$)
\begin{equation}
S = - \frac1{16 \pi G} \int_\mathcal{M} d^{d+1}x \sqrt{-g} \left( R + \frac{d (d-1)}{L^2} \right)
 + \frac1{8 \pi G} \int_{\partial \mathcal{M}} d^dx \sqrt{-\gamma} \left( - K + \frac{d - 1}{L} + \cdots \right),
\end{equation}
where $K$ is the trace of the extrinsic curvature of the boundary $\partial \mathcal{M}$.
For the SAdS black holes
\begin{equation}
ds^2 = - f(r) dt^2 + \frac{dr^2}{f(r)} + r^2 d\Omega_{d-1}^2, \qquad f = 1 - \frac{2 G_0 M}{r^{d-2}} + \frac{r^2}{L^2},
\end{equation}
the horizon of black holes is located at $2 G M = r_h^{d-2} (1 + r_h^2/L^2)$ and the associated temperature is
\begin{equation}
T = \frac{f'(r_h)}{4 \pi} = \frac{d - 2}{4 \pi r_h} + \frac{d r_h}{4 \pi L^2}
= \frac{d r_h^2 + (d - 2) L^2}{4 \pi r_h L^2}.
\end{equation}
There is a minimal value of temperature when $r_h = r_\mathrm{min} = \sqrt{\frac{d-2}{d}} \, L$:
\begin{equation}
T_\mathrm{min} = \frac{\sqrt{d (d - 2)}}{2 \pi L}.
\end{equation}
The free energy in the saddle point approximation is $F = - T \ln\mathcal{Z} \approx T \, S_\textrm{on shell}^{E}$.
The Euclidean on-shell action, $S_\textrm{on shell}^{E}$ for SAdS black hole and thermal AdS are obtained
by the integration outside the horizon in the full spacetime with $R = \frac{2 (d + 1)}{d - 1} \Lambda$:
\begin{eqnarray}
S_\mathrm{SAdS} &=& \frac{d}{8 \pi G L^2} \int_0^\beta d\tau \int d\Omega_{d-1} \int_{r_h}^{r_\infty} r^{d-1} dr
= \frac{\beta \Omega_{d-1} (r_\infty^d - r_h^d)}{8 \pi G L^2},
\\
S_\mathrm{AdS} &=& \frac{d}{8 \pi G L^2} \int_0^{\beta_0} d\tau \int d\Omega_{d-1} \int_0^{r_\infty} r^{d-1} dr
= \frac{\beta_0 \Omega_{d-1} r_\infty^d}{8 \pi G L^2}.
\end{eqnarray}
By matching the proper temperature $T(r) = T/\sqrt{- g_{tt}}$ on the horizon, we have
\begin{equation}
\beta \sqrt{f} = \beta_0 \sqrt{f_0} \quad \rightarrow \quad \beta_0 = \beta \sqrt{f/f_0}.
\end{equation}
Using this relation, we can obtain the difference in the free energies:
\begin{eqnarray}
\label{Eq_DF}
\Delta F &=& \frac{S_\mathrm{SAdS} - S_\mathrm{AdS}}{\beta} = \frac{\Omega_{d-1}}{8 \pi G L^2}
 \left( r_\infty^d - r_h^d - r_\infty^d \sqrt{\frac{f(r_\infty)}{f_0(r_\infty)}} \right)
\nonumber\\
&\approx& \frac{\Omega_{d-1}}{8 \pi G L^2} \left[ r_\infty^d - r_h^d - r_\infty^d \left( 1
 - \frac{G M L^2}{r_\infty^d} \right) \right] = - \frac{\Omega_{d-1} r_h^{d-2} (r_h^2 - L^2)}{16 \pi G L^2}.
\end{eqnarray}
The critical temperature for the HP transition corresponding to the point with zero free energy
difference $r_h = r_\mathrm{HP} = L > r_\mathrm{min}$ is
\begin{equation}
T_\mathrm{HP} = \frac{d- 1}{2 \pi L}.
\end{equation}
There is a first order phase transition, namely HP transition, indicating a phase change from
AdS thermal spacetime (higher free energy) to large SAdS black hole (lower free energy) for $T > T_\mathrm{HP}$.

Lets consider another approach to deriving the free energy via the black hole thermodynamics.
The related thermodynamic quantities, namely energy and entropy are given by
\begin{equation}
E = \frac{(d - 1) \Omega_{d-1}}{8 \pi} M, \qquad S = \frac{\Omega_{d-1} r_h^{d-1}}{4 G}.
\end{equation}
According to the first law
\begin{equation}
\delta E = T \delta S,
\end{equation}
the ``enthalpy''~\cite{Hu:2019lcy} $E$ is identical with the internal energy $U = E$.
Therefore the Helmholtz free energy (identical with the Gibbs function) for SAdS black holes is
\begin{equation}
F = U - T S = E - T S = - \frac{\Omega_{d-1} r_h^{d-2}}{16 \pi G L^2} ( r_h^2 - L^2 ).
\end{equation}
This is exactly the $\Delta F$ in~\eqref{Eq_DF}.

\end{appendix}

\end{document}